\documentclass[graybox]{svmult}

\usepackage{type1cm}        
\usepackage{makeidx}         
\usepackage{graphicx}        
\usepackage{multicol}        
\usepackage[bottom]{footmisc}

\usepackage{xcolor}
\usepackage[normalem]{ulem}

\usepackage{newtxtext}       %
\usepackage{newtxmath}       

\makeindex             

\begin{document}

\title*{Navigating Ethical AI Challenges in the Industrial Sector:\\ Balancing Innovation and Responsibility}
\author{Ruomu Tan and Martin W Hoffmann}
\institute{Dr. Ruomu Tan \at ABB AG, Corporate Research, Mannheim, Germany \email{ruomu.tan@de.abb.com}
\and Dr.-Ing Martin W Hoffmann \at ABB AG, Corporate Research, Mannheim, Germany \email{martin.w.hoffmann@de.abb.com}}
%
%
\titlerunning{Navigating Ethical AI Challenges in the Industrial Sector}
\maketitle


\abstract{The integration of artificial intelligence (AI) into the industrial sector has not only driven innovation but also expanded the ethical landscape, necessitating a reevaluation of principles governing technology and its applications and awareness in research and development of industrial AI solutions. This chapter explores how AI-empowered industrial innovation inherently intersects with ethics, as advancements in AI introduce new challenges related to transparency, accountability, and fairness. 
In the chapter, we then examine the ethical aspects of several examples of AI manifestation in industrial use cases and associated factors such as ethical practices in the research and development process and data sharing. With the progress of ethical industrial AI solutions, we emphasize the importance of embedding ethical principles into industrial AI systems and its potential to inspire technological breakthroughs and foster trust among stakeholders. This chapter also offers actionable insights to guide industrial research and development toward a future where AI serves as an enabler for ethical and responsible industrial progress as well as a more inclusive industrial ecosystem.
}

\section{Introduction: AI in the Industrial Sector}
\label{sec:intro}

Artificial intelligence (AI) is finding increasing application not only in the consumer world we are all familiar with --- from Copilot integrations into everyday computer and smartphone apps to home automation solutions and creative processes --- but also in the industrial sector. With industrial sector, we mean heavy industries like mining, oil and gas, steel production, and power plants, but the industrial sector also covers industries like cargo shipping, electronics manufacturing, water and wastewater treatment as well as power distribution.

\subsection{AI Roles of Industrial Companies}

\begin{figure}[t]
\includegraphics[width=1\linewidth]{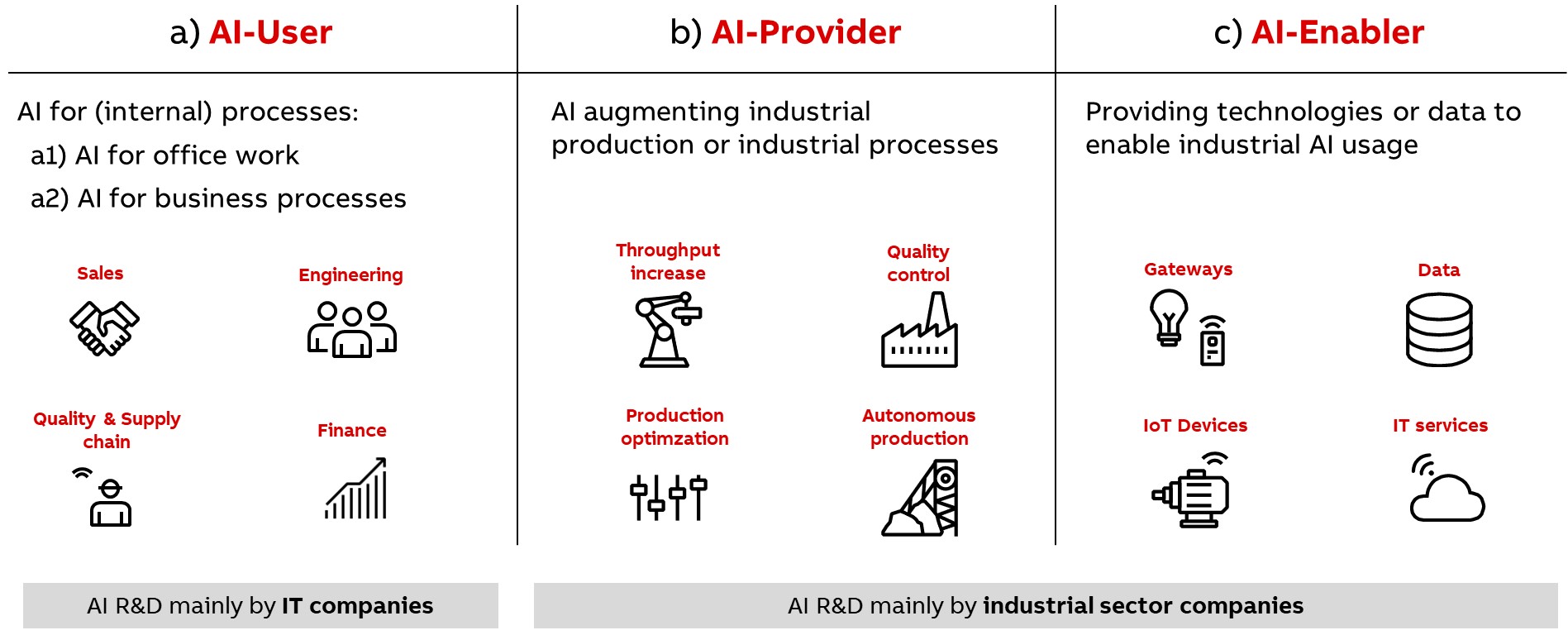}
\caption{Three roles an industrial company can take in the application of AI in the industrial sector.}
\label{fig:applicationFields}
\end{figure}

In the industrial sector, a company can take three different roles in applications of AI (cf. Figure \ref{fig:applicationFields}): a) Using AI for their own internal processes. The applications in this role can be split into a1) AI applications that are used by the people working in the industrial sector (e.g. office co-pilots, AI-based translation, ...) and a2) AI applications that enhance supporting business processes (e.g. procurement, human resources, customer relationship management, ...). b) Providing AI applications that augment industrial production or industrial processes (e.g. production process optimization, quality control, autonomous productions, ...), and c) providing technology to enable AI usage in the industry (e.g. communication infrastructure, IoT capable devices and systems, edge and cloud infrastructure or service, ...). Quite often, the roles b) and c) overlap, as the industrial systems are usually complex combined hardware \& software systems called ``Cyber Physical Systems (CPS)" \cite{hoffmann2021developing}. 

The AI applications used in role a1) are thereby covering roughly the same applications as any consumer can also use. Role a2) is also not specific to the industrial sector, but can be found similarly in any company type --- from insurances to logistics to retail. However, the technologies and applications in roles AI Provider and AI Enabler are specific to the industrial sector and technologies, and applications cannot be directly transferred from other business settings.

\subsection{Purpose \& Structure of this Chapter}

The purpose of this chapter is to provide a guidance to the reader how to navigate ethical AI challenges in the industrial sector to balance innovation and responsibility in industrial companies. 

\begin{figure}[t]
\includegraphics[width=1\linewidth]{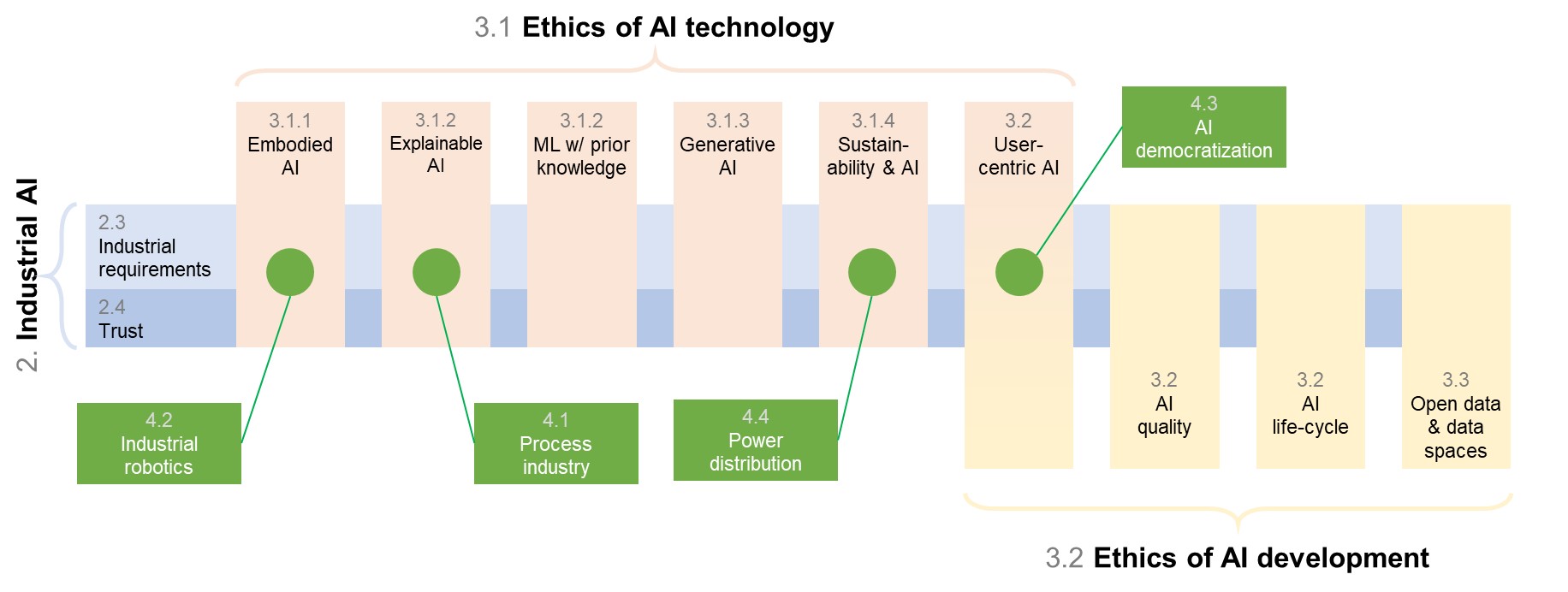}
\caption{Aspects and examples of the ethical usage of AI in the industrial sector. The numbers correspond to the sections in this chapter. Green boxes represent examples of industrial AI.}
\label{fig:structure} 
\end{figure}

This chapter will focus on AI applications and technologies in companies that take the roles of AI Provider and AI Enabler of the described taxonomy as these are the areas where industrial sector companies undertake their own research and development in AI. Ethical considerations of AI are explored in terms of three different aspects. i) Ethical considerations derived from and addressing requirements from different industrial roles (cf. Figure \ref{fig:applicationFields}), ii) ethical technological AI approaches, and iii) ethical methods integrated in the research and development (R\&D) processes of AI applications for the industrial sector. This chapter is structured along these three aspects of AI ethics. Figure \ref{fig:structure} visualizes the structure of the chapter.

\section{AI Ethics in the Industrial Context}
\label{sec:EthicsInd}

It is challenging to find commonly agreeable definitions of AI \cite{wang2019defining} and of ethics \cite{kazim2021high}. This makes it difficult to write down a definition of AI ethics. For this chapter we utilize the definition that (artificial) ``\textit{intelligence is the capacity of an information-processing system to adapt to its environment while operating with insufficient knowledge and resources}'' \cite{wang2019defining}. Ethics can be understood ``\textit{as the rational and systematic study of the standards of what is right and wrong}'' \cite{kazim2021high}, but is commonly mixed with morality and the theme of law \cite{kazim2021high}. We therefore define for this chapter ``AI ethics in the industrial context'' as

\begin{center}
``\textit{the understanding of the industrial sector of what is right and wrong in the development and usage of intelligent and adaptable industrial systems with laws as well as the morality and awareness of the developers and users as limits}''.
    
\end{center}

We have seen many different approaches to AI ethics in the industrial sector over the past decade. Most often, industrial companies put down their approaches as internal guidelines on AI ethics. The non-governmental organization ``Algorithm Watch" has compiled an extensive repository of more than 165 AI ethics guidelines over the past years\footnote{\url{https://inventory.algorithmwatch.org/}}, but recently terminated their efforts, due to decreased community support and less activities from the companies releasing new or revised guidelines. Most of the guidelines where released during the peak of Industry 4.0 between 2017 and 2019 to counter one of the main criticisms of collecting too much data and using it for unethical purposes. AI ethics initiatives in industrial companies ranged from manifests how to handle and utilize industrial data to the creation of AI ethics boards advising companies' executive committees. Besides companies trying to navigate the --- by that time --- unknown jungle of ethical challenges of industrial AI, some industrial sectors had, early on in the discussion of increasing digitalization, agreed on guidelines for the responsible sharing and usage of data, see e.g. the Association of the Electrical and Digital Industry (ZVEI) \cite{ZVEIguidelinesDataPlatforms}. Detailed analyses of AI ethics guidelines from the industrial sector (and other business sectors) can be found in the literature \cite{attard2023ethics,hagendorff2020ethics,jobin2019global}, which e.g. found that guidelines converge around ''five ethical principles: transparency, justice and fairness, non-maleficence, responsibility and privacy'' \cite{jobin2019global}. 
More recently, communities like the World Economic Forum (WEF)\footnote{\url{https://initiatives.weforum.org/ai-governance-alliance/}} have picked up the topic again and released recommendations and frameworks on responsible generative AI \cite{alliance2023presidio, alliance2024presidio} and best practices for investors in responsible AI \cite{alliance2024playbook}. These are however not focused exclusively on the industrial sector.

Besides differences in the implementation, all of these company- or community-driven efforts had a common goal: i) establishing trust in novel digital, data-driven and AI-enabled systems towards their (industrial) customers, ii)  manifesting an image of a trustworthy company towards society, and iii) preventing (unnecessary) legislative regulation by national and multi-national governments.

Over the past years, the discussion about industrial AI ethics has shifted to the legislative spectrum, with the creation of AI regulations in e.g. the EU (Data Act (Jan. 2024), AI Act (Aug. 2024, based on work of the High-Level Expert Group on AI\footnote{\url{https://digital-strategy.ec.europa.eu/en/library/ethics-guidelines-trustworthy-ai}}), Cyber Resilience Act (Dec. 2024)), Canada (Artificial Intelligence and Data Act (Apr. 2023)) or California (Generative Artificial Intelligence Law (Sep. 2024)). Readers interested in learning more about international AI regulations are pointed to online databases covering the topic\footnote{\url{https://www.whitecase.com/insight-our-thinking/ai-watch-global-regulatory-tracker}}.

\subsection{Demystifying ``Ethics" in Industrial Context}

Quite some efforts have been taken by the industrial sector to research and develop (R\&D) ethical AI solutions ever since datadriven solutions found their way into the industry. However, industrial players in most cases did not use the wording ``ethics" to describe their R\&D activities. The reason for this is likely multifold. For one, ``ethics" is most often perceived as countering business goals. Second, companies might not have been aware that their activities fall under such terminology. Thirdly, business ethical aspects are commonly handled by separate departments within companies. Especially large corporates usually maintain executive-level-governed teams on legal, integrity, and sustainability. These were in the past also the teams responsible for AI ethics guidelines described in this section.

\subsection{Requirements on AI Solutions in Industrial Manufacturing}
\label{sec:requirements}

The implementation of AI methods into information technology (IT) and operational technology (OT) systems in the industrial sector poses unique requirements compared to the consumer market \cite{Hoffmann2020}. In the consumer market, data is most often generated by humans, e.g. photographs of meals or expressions of political opinions on social media, and usually provides contextualized and frequently even multi-modal (i.e. image and descriptive text) information. Furthermore, there are large amounts of similar data available. In the industrial sector, the majority of the data is time-series or tabular data from machines or OT systems. Although there are a lot of machines of the same build, every machine is engineered to its specific purposes and therefore data across machines from the same type cannot directly be analyzed together. Furthermore, the machines are usually part of cyber-physical systems (CPS), where sensors, electronics, actuators, edge devices as well as computer clouds are interconnected. Table \ref{tab:requirements} provides an overview of industrial requirements and corresponding challenges of industrial AI.

\begin{table}[!t]
\caption{Requirements for industrial AI solutions, expanded from \cite{Hoffmann2020}.}
\label{tab:requirements}       
\begin{tabular}{p{2cm}p{3.5cm}p{5.8cm}}
\hline\noalign{\smallskip}
Area & Requirement & Addressed challenges \\
\noalign{\smallskip}\svhline\noalign{\smallskip}
Adaption & Stepwise introduction & Protection of previous investments \\
 & Human in the loop & Mistrust in new software tools \\
 & Data availability & Small, unbalanced, and unlabeled data\\
 &&\\
Engineering & Virtual learning & Software systems needs to operational from the point of commissioning \\
 & Adaptation & Every customer system is (slightly) different, and new AI needs to be easily adapted (or shall adapt itself)\\
 & Simplicity (hiding of complexity) & Industrial users are domain experts, not data scientists \\
 & Life cycle & Industrial hardware systems run for 10\textendash40 years. Software and AI systems are expected to have similar life cycles. \\
 &&\\
Embedding & Stacking of AI decisions & AI part of larger OT and IT systems, potentially also containing AI features\\
 & Trust space and borders & AI system needs to be limited to tested scope of influence\\
 & Knowledge distribution & Industrial production may never be depended only single point of knowledge, neither user nor vendor \\
 &&\\
Safety / Security & Safety & Safety is the top priority for any factory or plant \\
 & Robust against adversarial inputs & Malfunction of industrial systems can cause major economic up to environmental damage \\
 & On-premise computations & OT systems are generally isolated from IT systems and the internet  \\
 & Limited system scope & Generative AI introduces capability overhang which may infer security vulnerabilities or leakage of proprietary knowledge \\
 &&\\
Trust & Traceability and transparency of decisions & Legal compliance in critical infrastructure \\
 & Bias-free & Customers maintain a wide variety of machines and systems from different vendors \\
 & Confidence measure & Human user wants to understand how certain an AI model is \\
 & Quality classification & Not every prediction is 100 \% certain \\
 & Proof of capabilities & The correctness of predictive maintenance or control is difficult to prove, as faults ideally never happen, but it remains unclear if a fault would have happened \\
 & Deterministic output & Users expect to get deterministic output from software systems, generative AI introduces hallucinations \\
 && \\
 Business model & Fixed price & Industrial vendors and customers accustomed to CAPEX business models, not recurring services with variable running costs \\
  & Scalability & AI solutions only economical if rolled out to many customers with minimal engineering efforts \\  
  & Legal compliance & Globally varying AI laws, but multi-nationally operating industrial customers \\
\noalign{\smallskip}\hline\noalign{\smallskip}
\end{tabular}
\end{table}

This situation of smaller data-pools, most often also containing only sparse information, as well as specific requirements for cyber security, safety and engineering processes, lead to different methodological and technical approaches in developing, implementing and maintaining industrial AI solutions. CPS add further multi-disciplinary challenges in the development of industrial AI solutions, when physicists, engineers, and computer scientists needs to work together \cite{hoffmann2021developing}.

\subsection{Trust is All You Need?!} 
\label{section:trust}

The foundation of all above-listed requirements from industrial customers and users of AI is ``trust", which is also a major general challenge of AI \cite{lockey2021review}. In the industrial sector this is specifically crucial, as customer-vendor relationship are built and maintained over decades with hardware and software systems running reliably and deterministic over five to ten or even more years. AI now introduces a variability of output, uncertain life-time behavior and potentially variable (computing) costs. Furthermore, industrial users like plant operators or maintenance technicians are not knowledgeable with regard to AI or even software technologies and will turn off software systems at the first incident when these systems produce false and to them unexplainable output. Additionally, workers and workers' councils are fearing privacy and surveillance issues as well as replacement of humans by machines. This mistrust, along with general AI ethics problems, can strongly impact business success \cite{naik2022legal}.

\section{AI Manifestation in the Industrial Sector}

The rapid manifestation of AI in industrial use cases stems from adoption of the latest technologies and corresponding R\&D practices as well as efforts with regard to technological and organizational progresses; therefore, the ethical challenges in such manifestation also concern both the technological development and the organizational practice. In this section, we aim to highlight the practical technological and organizational trends in the industrial sector regarding the R\&D of ethical AI solutions. 

\subsection{Technological Trends}

AI manifestation in the industrial sector empowers use cases that were not possible in the past, extends the capability of humans working in an industrial setting, and helps to remove humans from harsh working environments. In this section, several examples of such AI manifestation are discussed with the focus on their implications for ethical aspects.

\subsubsection{Embodied AI: When AI Meets Reality} \label{embodied_ai}

Embodied AI refers to AI systems that are integrated in physical entities and systems, such as robots and vehicles, which have the ability to interact with the environment and learn from the interactions. Embodied AI symbolizes the convergence of AI and machines such that AI is enabled to interact directly with the physical world, sometimes in a human-like way. Capabilities of autonomous systems like robots are boosted with advancements in various research fields, such as computer vision, deep learning, control theory, and mechatronics engineering; these systems are able to perceive the real world, analyze and synthesize information, make decisions, and execute them to make an impact on the real world. 

Embodied AI starts to manifest itself in industrial use cases. Autonomous robots, as an example, can be found in various scenarios, including factory automation \cite{ren2024embodied} and maintenance in adversarial working conditions like sub-sea infrastructure \cite{saad2020advancing}. Moreover, with the current progress of AI deployment in industrial processes and their components, such as sensors, drives, control systems, and production scheduling systems, we foresee \textit{Embodied Industrial AI}, namely autonomous industrial systems that are smarter and more autonomous, to be a promising area of future research and application. Control, optimization, and maintenance of such autonomous systems will require less human intervention and fewer decisions from human experts, achieving overall higher levels of autonomy in production processes \cite{GAMER2020101}. The trend of Embodied Industrial AI will contribute to ethical practices in industry by reducing cognitive burdens of humans working with these systems, removing them from harsh or dangerous environments, minimizing the likelihood of human errors and thus improving safety and reliability, and addressing demographic challenges in the workforce. 

Ethical considerations of Embodied Industrial AI will be critical; one attempt to address them is to take inspiration from other domains of Embodied AI. As pioneers of materializing Embodied AI systems at large scale in real life, the autonomous vehicle industry has spearheaded identifying ethical challenges and addressing these challenges from technical and other perspectives \cite{MARTINHO2021556}. We believe that many of the ethical challenges identified for autonomous vehicles, such as safety and reliability, accountability, human control, and auditing, may be transferrable to AI embodiment in autonomous industrial systems whilst new ethical challenges will emerge as Embodied Industrial AI is integrated into the industrial environment.

\subsubsection{We Built the Machines, So Why is AI Still a Black Box?}

In the industrial sector, the vast majority of data for AI solutions is produced by machines as well as OT and IT systems. This brings up the question why one should employ black-box data-driven machine learning (ML) methods instead of analytics and physics-based methods. There are multiple reasons for this: i) It is not economical to create a detailed physics-based digital twin of every machine, ii) production systems are too complex for real-time physics-based simulations, and iii) factories and plants are overhauled every couple of years, but engineering data is not updated. Nevertheless, industrial users, who are domain experts for the machines and industrial systems, are reluctant to accept black-box support systems. 

To foster acceptance and economical usage of digital twins in the industrial sector, explainable AI (XAI) is one of the major R\&D directions \cite{kotriwala2021xai}. XAI sheds light on the decision process of AI systems, makes them more transparent and traceable, and can thereby support user acceptance of the systems. This technology is maturing quickly, and will likely be included in most upcoming AI solutions for the industrial sector, see also Section \ref{section:XAIexample} for examples.

Another promising R\&D direction is the combination of data-driven and physics-based methods, so-called machine learning with prior knowledge \cite{watson2024machine}. This approach combines prior knowledge of the machines and systems with adaptive machine learning methods. It thereby can compensate sparsity of dynamic data in the industrial sector with static knowledge. The accompanying AI methods are currently in academic research, and a wide-scale adaptation in the industrial sector will likely take a longer time.

\subsubsection{Generative AI: New Ethical Landscape}

Generative AI refers to AI systems that can generate new content which is similar to that generated by human intelligence. It is different from analytical AI where specialized models are trained to derive insights from a specific collection of data. Pre-training of generative AI models with huge amounts of data not only enables the generation of new content but also results in better reasoning and decision-making abilities, reflecting higher levels of intelligence in AI systems.

Unethical development and usage of generative AI lead to societal risks, such as privacy infringement, bias, deep fake, and other misinformation. Furthermore, it is worthwhile to discuss the ethical implication of using output from professionals to train an AI model, which subsequently potentially replaces the jobs just these professionals. Many research activities are centered around alignment and safety when training and applying these generative AI models \cite{openai2024safetyupdate, openai2024disrupting}. Nevertheless, controversy exists regarding whether the current effort put into these research topics is sufficient due to the heterogeneous capabilities of generative AI models and their stochasticity.

In the industrial sector, the development and deployment of generative AI-based solutions have recently sparked across a large variety of use cases. Since the exploration is still in its early phase, ethical and responsible development and usage of generative AI for industrial use cases remains an open topic. As academic researchers also pointed out, the traditional landscape of engineering ethics should be extended to account for the introduction of generative AI \cite{DANIEL2024100168}. This is generally true in various fields of application of (generative) AI; nevertheless, unethical and misused AI in industrial settings, e.g. chemical plants, may have substantial consequences for the safety and well-being of humans, critical infrastructure, and the environment and should therefore be handled with extra caution and awareness. 

In addition, the introduction of generative AI will enable new functionalities and features in industrial use cases, resulting in an increased scope of impact of AI-based solutions and hence an extended landscape of application of engineering ethics. For example, novel generative AI-based operator support systems in process plants can answer operators' questions and provide recommendations for handling certain operating scenarios, which will expand operator capabilities and reduce response time to emergencies. Such expansion, however, also creates spaces for unethical usage: an operator could query the system to get instructions on how to bring the process into an unsafe operating condition that appears to be more profitable. Such instructions are not part of the standard operating procedures, but generative AI may still give recommendations based on its prior knowledge of control and operation of similar industrial processes. Following the recommendation could lead to unsustainable and unsafe production and even cause major process upsets.

The ethical challenge here applies to both the design of the generative AI-based support system and the awareness around ethical usage and practice among its users. A new engineering ethics landscape both in R\&D and practice should be a key consideration as generative AI becomes more accessible in the industrial sector.

\subsubsection{AI’s Path to Sustainability Amid High Energy Demands}

The main environment-ethical questions around all AI developments is most often how much energy AI models shall use, both during model training, as well during model inference \cite{gitzel2024estimating, desislavov2021compute,kar2022can}. 
Model training thereby usually consumes much more energy than a single inference. However, inference occurs orders of magnitude more often, especially if the model is frequently used or long-lived. In the ideal case, the usage of an AI model saves more resources (including energy, materials, and others) \(E_{saved}\), e.g. through AI-driven production optimization, than energy was used during its training \(E_{training}\) and inferences \(E_{inference}\) during the lifetime of the model.

\[ E_{saved} > E_{training} + \sum_{lifetime}^{} E_{inference} \]

In the industrial sector, this is not purely a sustainability question but also very plainly an economical question, i.e. if the business value \(C_{revenue}\) of an AI model outweighs the costs for energy and man-efforts to develop \(C_{training}\) and run \(C_{inference}\) an AI model within a reasonable time, a so-called ROI duration (return on invest duration).

\[ C_{revenue} > C_{training} + \sum_{ROI}^{} C_{inference} \]

Furthermore, in the industrial sector specifically, there are hardware constraints for inference, and to a lesser degree also the modeltraining, as AI models are being deployed on low-power industry PCs (most often called ``edge devices") or even industrial controllers \cite{meuser2024revisiting}. 

A first step to assess this topic is to measure the energy consumption of AI models through simple adaptation of the engineering workflow, as described in \cite{gitzel2024estimating}. Then, the AI model architecture can be adapted to suit pre-given energy consumption requirements. Alternatively, small AI model architectures, like TinyML \cite{kallimani2024tinyml}, can be used to deploy the inference of models on resource-constrained devices like edge PCs or controllers \cite{banbury2021micronets,huang2022real,immonen2022tiny}

In order to responsibly handle new industrial AI applications, we recommend to
\begin{enumerate}
  \item consider the target hardware for model inference before starting to develop industrial AI models,
  \item test and measure energy consumption during model development, training, and inference, and
  \item minimize AI model size for each given task.
\end{enumerate}

Such approaches will both save natural resources as well as maximize the business case of the AI applications.

\subsection{Ethical Development of Industrial AI Solutions}
\label{sec:devMethods}

Besides developing ethical AI solution for the industrial sector, it is also of importance to look at how these solutions are developed develop these. For the development process, the ethics focus lies on building up trust between the solution provider and users (cf. Section~\ref{section:trust}, e.g. by making the R\&D process transparent, as well as creating trust in the AI solution itself, e.g. by introducing quality measures into the R\&D process. Furthermore, the usage and sharing of open data (cf. FAIR data in Section~\ref{sec:openData}) plays a role in the ethical development of industrial AI solutions. The concepts of ``Ethics by Design'' \cite{dignum2018ethics} should also be considered in some way in the R\&D process, similar to former discussions in the robotics sector \cite{dodig2012robots}.

The requirements from the industrial sector for AI solution (cf. Section \ref{sec:requirements}) lead to a need for a structured, documented, and value-focused development process for AI applications. Such process has long been developed and refined for data-intensive industrial applications (CRISP-DM) \cite{wirth2000crisp}. The requirements changed when machine learning methods became more complex over the past two decades. Customers as well as developers disliked the iterative nature of CRISP-DM, as development cycle length and costs seemed nondeterministic. This led to the creation of linear, or at least linear-looking, AI development processes specific to the industrial sector, e.g. \cite{kloepper2020a, HasterokStompe2022}. These processes also accounted for providing value to the user in every development step \cite{kloepper2020a}, which vastly increases the users' trust and customers' willingness to share  data for. 
Further development processes introduce quality measures in the developed industrial AI solutions, such as a quality-focused CRISP-DM \cite{studer2020} or the ETAMI framework \cite{etami2023}. 
All processes have in common that they put industrial user into focus to ensure long-term trust in the AI solutions, so-called user-centric co-creation \cite{russkamp4involving, kloepper2020a}.


There have also been initiatives to create certification measures for the quality, life cycle, trustworthiness, and ethical development of AI solutions, e.g. \cite{winter2021trustedartificialintelligencecertification, PLSzertifizierung, Peylo_Slama_Hallensleben_Hauschke_Hildebrandt_2022}. These initiatives did not find wide adoption in the industrial sector, mainly due to the large administrative overhead of the proposed certification processes as well as new legislative requirements in the EU which partly covers these aspects (cf. Section~\ref{sec:EthicsInd}). Nevertheless, it is crucial for an ethical development of industrial AI to have measure to ensure quality of and trust in industrial AI solutions. Such measures should include company-internal trainings for engineers on AI technologiess, R\&D processes as well as ethical aspects.

In summary, the ethical development to foster long-term trust in and quality of industrial AI should:
\begin{enumerate}
  \item follow a structured, transparent and value-bringing development approach,
  \item put the user into the focus of the development and co-create with the user together,
  \item take the long-term quality of ML solutions already in the development phase into consideration, and
  \item be run by well-trained engineers, not only in machine learning but also in areas such as ethics and sustainable ML, who are aware of the industrial requirements for AI systems.
\end{enumerate}


\subsection{Ethical Data Sharing and Industrial Business Interests}
\label{sec:openData}

As in the consumer sector, data is also in the industrial sector a crucial prerequisite for the development of AI solutions. In an ideal ethical world, data should be shared along the value-chains, within value-networks or even across complete market segments and also with academic researchers and the public. This would create larger and information-rich data pools to develop more robust and stable AI solutions. Specific to the industrial sector, the data needs to be labeled, i.e. annotated by domain experts (cf. Section \ref{sec:democratization}), and needs to be information-rich, i.e. cover a high number of ``interesting situations" like anomalous machine behavior or plant shutdowns. Ideally, all data should also follow the FAIR principle \cite{wilkinson2016fair} of findability, accessibility, interoperability, and reusability. Findability e.g. reflects that datasets need to be stored in suitable repositories and need to include a meaningful description of the dataset in its meta data. Accessibility e.g. refers to the situation that access to dataset shall not be restricted by unnecessary technical hurdles or be limited to certain user groups. Interoperability e.g. suggests to use estabslihed or standardized data formats. Reusability e.g. addresses the issue that legal hurdles may prohibit the reusage of available data. Although there is a clear need for the industry to go FAIR \cite{van2020need}, a generalized, ethical sharing of data from and within the industrial sector stands clearly in conflict with the business interest of the companies in this sector, on both the vendor and the customer side. Industrial data contains business-critical information about vendor structures, product and production know-how, as well financial data and personal data.

Currently, data in the industrial sector is being shared on a contractual basis bilaterally between an AI vendor and an AI customer, or multilaterally in industrial consortia or public private partnerships (PPP), e.g. in public-funded consortium research projects like \cite{hoffmann2020integration}. Examples for industrial consortia with structured data sharing are CATENA-X for the automotive sector \cite{schoppenthau2023building,mugge2023empowering} and GAIA-X for the manufacturing sector \cite{braud2021road}. Such approaches ensure contractual security of business interests and keeping control of the data, whilst still fostering increased data accessibility and AI innovation.

For the development of industrial AI solutions (cf. Section \ref{sec:devMethods}) repositories of open data from and for industrial applications can be used; examples include the TUDataset database for graph-ML applications \cite{morris2020tudataset}, the NASA Prognostics Center of Excellence's Data Set Repository\footnote{\url{https://www.nasa.gov/intelligent-systems-division/discovery-and-systems-health/pcoe/pcoe-data-set-repository/}}, the EU Data Portal\footnote{\url{https://data.europa.eu/en}}, and the Open Power System Data Repository\footnote{\url{https://open-power-system-data.org/}}.

In order to increase the value of industrial data, there needs to be agreement on standardized data formats and especially information models, such as the asset administration shell \cite{tantik2017integrated}. These are also crucial for long data life cycles in plants \cite{wagner2017role} as well as standards for handling data of digital twins as part of CPS \footnote{\url{https://industrialdigitaltwin.org/}}.

As there is great economic potential of machine-generated, non-personal data in the EU \cite{DeloitteLLP2018}, similar to all other parts of the world, the EU follows the idea of a ``data union" with data sovereignty. In accordance with the new legislation in the EU, cf. Section \ref{sec:EthicsInd}, the EU supports the establishing of industrial data spaces like GAIA-X or CATENA-X. Furthermore, national authorities are discussing new laws regarding mandatory data sharing with academic research.

\section{Examples of Ethical Industrial AI Solutions}

In this section, we further examine several examples of industrial AI solutions that actually contribute to the ethical aspects of the corresponding industrial use cases. With these examples, we aim to highlight that AI is not only an additional aspect to be considered when addressing ethical challenges in the industrial sector but also an instrument to help facilitate ethical practices and that ethical challenges can bring opportunities for research and development for industrial AI.

\subsection{AI You Can Trust}
\label{section:XAIexample}

Section \ref{section:trust} discussed how trust is fundamental to the ethical usage of AI in the industrial sector. To cultivate trust in human users, stakeholders, and regulators, XAI fosters transparency, trust, and accountability in AI-driven decision-making. XAI makes it easier for human, particularly non AI-experts, to understand how AI systems work, examine the influential factors on the output of their AI systems, implement changes in the behavior of their AI systems, and become aware of the boundaries and limitations of the AI systems.

Researchers have started addressing the explainability of AI-based solutions in industrial use cases including operator support, predictive maintenance, and alarm rationalization \cite{manca2023xaiprocesslens,manca2023explainable}. For example, \cite{manca2023xaiprocesslens} proposed a dashboard for process experts to analyze the various phases of a batch reactor by training clustering models using time series data from multiple batches and highlighting the process variables that are most representative of certain phases. Explainable dashboards are particularly relevant when operators face with numerous information flows enabled by AI.

Moreover, \cite{kotriwala2021xai} provided an overview of XAI in the process industry, illustrating that explainability reduces the black-box nature of AI-based solutions adopted in industrial use cases in order to address the particular risk awareness in industry and provide justification for decision-making. The paper also outlined potential challenges and research directions, such as the need to focus on domain experts and domain know-how, the importance of dynamic and adaptive explanation as well as always keeping humans in the driver's seat. Incorporating XAI into industrial applications is not just a technical advancement but an ethical necessity.

\subsection{Human and Machine Collaborate, Safely}

In Section \ref{embodied_ai} we discussed the trend of Embodied Industrial AI, where AI and mechanical objects converge and collaborate to address industrial use cases. In fact, the convergence of AI and mechanical objects will not only enable the objects to execute sophisticated tasks in the real world but also enhance the ethical execution of these tasks when compared to traditional objects. The adoption of AI can achieve better perception, smarter decision-makin,g and more accurate actuation in these mechanical systems, thus better addressing ethical challenges, such as the safety and well-being of humans, in an industrial environment.  

One solid example is the industrial collaborative robots that can plan their movement in order to reduce the risk of collision with humans in their close proximity. Machine learning algorithms are developed for the robots to distinguish ``interactions'' with humans, which are normal and should be facilitated, from ``collisions'' that can be dangerous to human beings and should be avoided. Once a potential collision is identified, the robot controller reduces the speed of the robot and disengages to prevent applying excessive force on human beings \cite{Kerestedjian2019using}. Moreover, it is possible to optimize the trajectory and speed of robot movements with the constraints of power and force in order to limit potential collisions \cite{stuhlenmiller2024optimal}. Then robots are capable of moving in a time-efficient path for executing their tasks whilst guaranteeing any potential collisions with other objects, particularly humans, that may appear on their paths, will be within the safety limit.

In this example, the ethical challenge of human physical safety was first identified; novel, AI-enabled solutions were then developed to address the challenge, resulting in both safer and more ethical human-robot collaboration and advancement in research and application of various AI techniques. Hence, we believe that ethical challenges and requirements should not be hurdles for industrial AI application, rather opportunities for innovative research and new technology development. Such exercises will become more relevant as Embodied Industrial AI continues to be integrated in various industrial systems and processes.

\subsection{AI as the Great Equalizer}
\label{sec:democratization}

The industrial sector has historically been characterized by significant barriers to entry, with access to cutting-edge tools, expertise, and resources often unevenly distributed. These challenges also hold when individuals choose to start their careers in the industrial sector.

On the other hand, AI can make advanced tools, technologies and domain knowledge more accessible to a broader group. For example people can communicate with large language models using natural language to harness their power to carry out tasks in an industrial setting, such as generating control codes \cite{koziolek2024chatgpt}, decision support for operators \cite{tan2024operator}, and advanced data analysis, that would conventionally require expert-level know-how and specialized tools. This 
break down of traditional barriers creates a more inclusive environment for people with diverse backgrounds to engage with and contribute to industrial use cases as well as to collaborate with others.

Particularly when AI systems are deployed in the industrial sector, the users can often be domain experts who posses valuable know-how of industrial use cases but are less experienced about AI. Mechanisms like active learning are designed to enable domain experts to provide inputs and feedback to improve the performance of industrial AI systems \cite{ahmad2022active}. Moreover, by leveraging technologies like generative AI \cite{tan2024operator}, AutoML \cite{kurian2021boat}, and user-friendly AI dashboards \cite{manca2023xaiprocesslens}, domain experts are empowered to carry out analysis of their data, optimize the performance and processes, and even create novel and customized features that benefit their work the most.

By making complex technologies accessible, AI can transform industries into more inclusive ecosystems, ensuring that innovation is no longer limited to a select few but shared by all, but it also brings up questions around ownership of ideas, patentability and trade secrets.

\subsection{Embracing Sustainability Megatrends}

With AI or not, megatrends like sustainability, social equity and advanced healthcare, have been gaining momentum, driven by societal demands and global challenges. These megatrends exist as guiding principles for industries, and researchers have been innovating to advance them. Now with the development of AI, these ethical megatrends could gain a transformative edge.

AI empowers industries to operate more sustainably by optimizing resource utilization, minimizing waste, and reducing environmental impact. For instance, AI-driven monitoring and prognostic systems can improve the reliability and the efficiency of grid components, enhancing the stability of power grids when renewable energy is introduced. Such systems are a key enabler for large-scale usage of renewable energy in the industrial sector \cite{hoffmann2020integration,gitzel2024towards}. In general, AI-based methods are currently driving energy efficiency, resource utilization optimization, and environmental impact reduction across a wide range of industrial processes. 

There are more ethical trends that are facilitated by AI-based solutions, such as AI-based monitoring that can support responsible and ethical supply chain; AI-driven liberalization of tools and knowledge is creating a more inclusive and equal industrial ecosystem.

\section{Conclusions}

AI is not only an additional aspect to be considered when addressing ethical challenges in the industrial sector but also an instrument to help facilitate ethical practice in industry. In summary, we hope to have shown to the reader that

\begin{enumerate}
    \item ethical considerations are important for successfully establishing industrial AI applications,
    \item AI technology development is currently faster than AI ethics development,
    \item many R\&D activities address ethical industrial AI, without naming it so explicitly, 
    \item ethics in industrial AI opens new research space for AI technologies,
    \item ethics in AI can be seen as an enabler for better adoption of AI solutions in industrial manufacturing, instead of being perceived as burden, and
    \item a lot remains to be done --- in terms of technology adaptation, R\&D practices, and organizational adaptation.
\end{enumerate}

General challenges in the adaptation of ethical AI in the industrial sector include FAIR data sharing often being in conflict with protecting business value and ``ethics"-washing being a risk for industrial companies who need to stay trustworthy.

We recommend the industrial sector to 
\begin{itemize}
    \item cautiously embody AI into machines and industrial systems,
    \item add explainers and physics-based components to industrial AI solutions,
    \item adapt AI for sustainability applications,
    \item responsibly integrate generative AI,
    \item make use of the computationally smallest viable machine learning models to reduce energy consumption (and thereby cost),
    \item consider the long life cycles of industrial systems early in the AI R\&D process,
    \item utilize transparent, structured, and user-centered R\&D approaches,
    \item contribute to data sharing wherever economically possible, and
    \item consider adapting the wording of ethical AI in a responsible manner.
\end{itemize}

\section*{Acknowledgments}
The work on this chapter was partially funded by the German Federal Ministry of Education and Research (BMBF) under the grant number 01IS22030 (EXPLAIN project).

The authors would like to thank their colleagues Marie-Helene Westholm-Knebel and Ralf Gitzel for the thorough review of the draft and providing valuable inputs.

\section*{Transparency note}

The authors work for ABB AG, Germany, a company providing automation and electrical solutions to the industrial sector.

The findings and opinions expressed in this chapter are based on the authors' experience in R\&D of industrial AI and do not necessarily represent the opinion of their employer.

\bibliographystyle{ieeetr}
\bibliography{industrialAIethics}

\end{document}